# Thermal Properties of the Hybrid Graphene-Metal Nano-Micro-Composites: Applications in Thermal Interface Materials

Vivek Goyal[1,2] and Alexander A. Balandin[1,*]

[1]Nano-Device Laboratory, Department of Electrical Engineering and Materials Science and Engineering Program, University of California, Riverside, California 92521 USA

[2]Texas Instruments, Dallas, Texas 75243 USA

## Abstract

The authors report on synthesis and thermal properties of the electrically-conductive thermal interface materials with the hybrid graphene-metal particle fillers. The thermal conductivity of resulting composites was increased by ~500% in a temperature range from 300 K to 400 K at a small graphene loading fraction of 5-vol.-%. The unusually strong enhancement of thermal properties was attributed to the high intrinsic thermal conductivity of graphene, strong graphene coupling to matrix materials and the large range of the length-scale – from nanometers to micrometers – of the graphene and silver particle fillers. The obtained results are important for thermal management of advanced electronics and optoelectronics.

[*] Corresponding author; electronic address: balandin@ee.ucr.edu ; group web-site: http://ndl.ee.ucr.edu

The continuous device down-scaling and increasing power densities have led to escalating hot-spot temperatures [1-3]. Thermal management became a critical issue for the next generation of electronics. The latter motivates development of novel thermal interface materials (TIMs) with enhanced thermal conductivity $K$. TIMs facilitate heat transfer across interfaces by reducing the thermal resistance between the heat-generating chips and heat sinks. The selection of suitable TIMs to fill the interface between a chip and a heat spreader is critical to the performance and reliability of various devices [4-7]. Both types of TIMs – electrically insulating and conducting – are needed for electronic applications [6-9]. The electrically conductive TIMs provide heat removal and electrical connectivity in computers and communication hardware. It is known that even a modest increase in $K$ of TIMs will translate into lower junction temperature and improvements in device performance.

Conventional TIMs use polymers or greases as the matrix, i.e. base, material and utilize large loading volume fractions $f$ (of up to ~70 %) of the filler, e.g. silver or silica particles, to achieve the thermal conductivity of 1 – 5 W/mK at room temperature (RT) [8-9]. The $K$ enhancement via loading of the metal particles is often limited by the thermal contact resistance $R_C$ of the particles. Large $f$ leads to problems with TIM's viscosity and other parameters defining their applicability as well as increases cost. Thus, there are strong motivations to create more efficient fillers, which would strongly increase $K$ of the resulting TIMs at low loading $f$.

Here we investigate thermal properties of the composites with the novel hybrid graphene – metal particle fillers, which can be used as efficient TIMs. We observed an unusually large increase in $K$ of a composite with a nearly constant electrical resistivity $\rho$ at a small loading $f$ over a wide temperature $T$ range. The high $K$ values and their weak $T$ dependence were explained by the unique physics of phonon transport in graphene and specifics of thermal properties of the hybrid fillers.

It was discovered that graphene exhibits superior thermal conductivity [10]. The intrinsic $K$ of sufficiently large graphene flakes can exceed that of bulk graphite along the basal planes, which by itself is very high: $K \approx 2000$ W/mK at RT [10-12]. The latter was explained the unusual features of the phonon heat conduction in two-dimensional systems [11-12]. Few-layer graphene (FLG) also has excellent thermal properties [13], which can be retained when FLG is incorporated within matrix materials [14]. Recent availability of inexpensive commercially produced graphene-FLG solutions allowed us to investigate hybrid fillers for TIMs where graphene flakes are combined with metal particles to produce an efficient heat conducting network.

For this study, we used electrically conductive silver epoxy with the silver particles as the filler. The sizes of the silver particles were in μm range. The graphene-FLG solution was prepared by the flake isolation via the density gradient ultra-centrifugation (DGU) [15]. The aqueous-solution-phase approach was enabled by the sodium cholate (SC) surfactant [16]. In DGU, the solution of graphite flakes is centrifuged to control the FLG thickness. To remove the thick graphite material from the dispersion, the centrifugation was performed at 15 K-rpm for 5 minutes. The final product was SC-encapsulated graphene-FLG sheets with the concentration of ~0.05 mg/mL. The utilized liquid graphene-FLG solution had the single layer graphene (SLG) and bilayer graphene (BLG) content of ~27% and 48%, respectively. We treated the solution thermally to reduce the surfactant concentration and improve the flake-to-flake contact.

To prepare the hybrid graphene-metal-epoxy composites, we dispersed the graphene solution in the silver epoxy, and applied the high-shear mixing followed by ultrasonication (Fisher Scientific FB-705). The epoxy hardener was added to the homogeneous mixture of the graphene-silver epoxy and shear mixed. Figure 1 (a) shows the scanning electron microscopy (SEM) image of the resulting hybrid graphene-silver-epoxy composite. One can see the silver particles as the μm-scale grains. The graphene-FLG flakes with nm-scale thickness and lateral sizes varying from nm to μm scale are more difficult to distinguish. Their presence and concentration were verified with the electron dispersive spectroscopy (EDS). The EDS data presented in Figure 1 (b) reveals carbon (C) atoms present together with silver (Ag) and oxygen (O) atoms. The left inset to Figure 1 (b) is the transmission electron microscopy (TEM) image (FEI-PHILIPS CM300) of FLG flakes utilized in the composite. The selected-area electron-diffraction patterns (SAED), collected during TEM characterization, indicated that graphene flakes retained their good crystal structure after all processing steps.

The homogeneous mixture of the epoxy with the nano-micro-fillers of graphene-FLG and silver particles was loaded into the custom-made disk-shaped stainless steel mold for curing at RT for 24 hours. The residual solvent was removed by baking the mixture at $120^0$C for 10 minutes. The samples were then heated at $80^0$C for 15 hours to remove remaining solvent and possible air bubbles. The right inset to Figure 1 (b) shows the molded composite disk. The samples were prepared with the graphene-FLG loading fraction varying from 0.5 to 3 weight %. Following the same procedure, we also prepared several reference samples replacing graphene-FLG filler with the commercial carbon black (Asbury Graphite Mills).

The thermal conductivity of the obtained composites was measured using the transient planar source (TPS) technique (Hot Disk TPS2500) [17-18]. This technique has been previously used for investigation of thermal properties of other TIMs such as the phase change materials [19] and industrial thermal greases [20]. For these measurements we sandwiched an electrically insulated flat nickel sensor with the radius 2.001 mm between two identical samples of the same composition. The sensor acted as the heat source and $T$ monitor simultaneously. The surfaces of the specimens were flattened and cleaned to reduce $R_C$ at the sensor-sample surfaces. The details of the TPS measurement procedures were reported by us elsewhere [21].

Figure 2 presents the measured $K$ of the commercial silver epoxy, the hybrid composites with different graphene-FLG loading and reference composites with the carbon black. The error bars represent the data scatter for several samples. More than ten samples were investigated for each data point to ensure reproducibility. The measured $K$ for the commercial silver-epoxy TIM was 1.67 W/mK, which is in agreement with the value provided by the vendor ($K \approx 1.6$ W/mK). The thermal conductivity of the reference composite does not increase noticeably with the addition of carbon black in the examined $f$ range. Carbon black is an amorphous material with low $K$ [14]. However, one can see a drastic $K$ increase in the composites with addition of the graphene-FLG filler. The RT thermal conductivity of TIMs with the graphene-FLG nano-micro-filler particles reaches ~9.9 W/mK at the small 5-vol.-% of the graphene-FLG loading. This is significant increase by a factor of ~6, which corresponds to ~500% $K$ enhancement as compared to the commercial silver-epoxy TIM.

The weight fraction calculation included graphene-FLG, epoxy resin and hardener, and SC surfactant and iodixanol used in the purification process. The weight fraction was converted to the volume fraction using the equation [22] $V_{GF} = W_{GF}/[W_{GF} + (\rho_{GF}/\rho_B)(1-W_{GF})]$, where, $V_{GF}$ and $W_{GF}$ are the volume and weight fraction of graphene-FLG flakes, $\rho_{GF}$ and $\rho_B$ are the densities of the graphene-FLG filler and base material, respectively. In this case, the base material is the silver epoxy with $\rho_B \approx 4$ g/cm$^3$. The volume of the graphene unit cell is equal to $V_0 = 3\sqrt{3}a^2 h$, where $a = 0.142$ nm and thickness of grapheme layer is $h = 0.35$ nm. Since the unit cell consist of 2 carbon atoms the mass of the unit cell is $m = 2M_{carbon}$. The graphene density can be estimated as $\rho_{GF} = m/V_0 = 2M_{carbon}/V_0 \approx 2.2$ g/cm$^3$.

Temperature dependence of $K$ is important from both physics and applications points of view. Figure 3 shows $K$ as a function of $T$ for our composites with $f$=1%, 3%, and 5% volume fraction of graphene-FLG. One can see that $K$ only weakly depends on $T$, increasing slightly with $T$ varying from 5°C to 75°C. The $T$ range was selected to correspond to the operating $T$ of the commercial silver epoxy. The weak $K(T)$ dependence is an extra benefit for TIM applications.

The physics of thermal transport in the investigated composites is complicated. The heat in graphene is mostly transferred by acoustic phonons while in silver particles it is transferred by electrons [14]. Composites are disordered systems and many factors may affect their $K$. In graphene as well as other crystalline materials $K$ decreases with increasing $T$ owing to stronger phonon Umklapp scattering [11, 14]. In bulk crystals $K$ is proportional to $1/T$. From the other side, in the disordered and nanostructured materials $K(T)$ can grow with $T$ as a result of better phonon transmission through the interfaces and decreasing Kapitza resistance at higher $T$ [23]. The interplay of these trends can lead to the observed nearly flat $K(T)$ in the relevant $T$ range. It was recently shown computationally that owing to its flat geometry and better coupling to the matrix material due to smaller Kapitza resistance, FLG can increase $K$ of the composites of up to two orders of magnitude at $f$=5% if the graphene layers are ordered [24]. This can explain the measured enhancement values in our composites where graphene-FLG fillers were random. A combination of different sizes of the filler particles extending from nm to mm are also beneficial for increasing $K$ via formation of the better percolation network [25-26].

The epoxy matrix is a weak conductor of heat and electrical insulator. However, the high loading fraction of silver particles in the pristine silver epoxy ensures that TIM is electrically conductive. Our measurements (Signatone/HP4142) of the silver epoxy revealed the electrical resistivity $\rho_e \approx 10^{-4}$ Ω-m. The low $\rho_e$ means that the silver particles touch each other forming an electrically percolating network. It is interesting to note that $\rho_e$ of our samples did not change noticeably with addition of graphene. The latter can be explained by the fact that since the composite is highly electrically conductive by itself the addition of only 5 vol.-% of graphene-FLG cannot change its $\rho$ substantially.

In conclusion, we observed an extremely strong enhancement of $K$ in the hybrid graphene-metal particles composites. The physical processes leading to such enhancement were explained qualitatively. The obtain results can lead to development of a new generation of the high-efficiency TIMs for thermal management of advanced electronics.

*Acknowledgements*

This work was supported, in part, by the Office of Naval Research (ONR) through award N00014-10-1-0224 on graphene heat spreaders, by the Semiconductor Research Corporation (SRC) and Defense Advanced Research Project Agency (DARPA) through FCRP Center on Functional Engineered Nano Architectonics (FENA). The authors acknowledge useful discussions on TIMs with Intel Corporation engineers, and thank the former members of the Nano-Device Laboratory (NDL) – Dr. S. Ghosh (Intel) and Dr. D. Teweldebrhan (Intel) for their help with measurements.

**FIGURE CAPTION**

**Figure 1:** (a) SEM image of the graphene-silver-epoxy composite. (b) EDS data for the composites indicating the presence and concentration of graphene-FLG particles together with the silver particles. Left inset shows TEM image of a representative graphene flake used in the composites. The scale bar is 25 nm. Right inset shows an optical image of TIM sample used in thermal measurements. The diameter of the sample is 10 mm and its thickness is 1 mm.

**Figure 2**: Thermal conductivity of the pristine silver epoxy, hybrid graphene-FLG-silver-epoxy composites and the reference silver epoxy-carbon black composites as a function of the volume fraction $f$ of the graphene-FLG nano-micro-filler.

**Figure 3**: Thermal conductivity of the hybrid graphene-FLG-silver-epoxy composite as a function of temperature for 1%, 3% and 5% of the volume fraction of graphene-FLG filler loading. Note that the thermal conductivity almost does not change in the examined range, which is important for TIM applications.

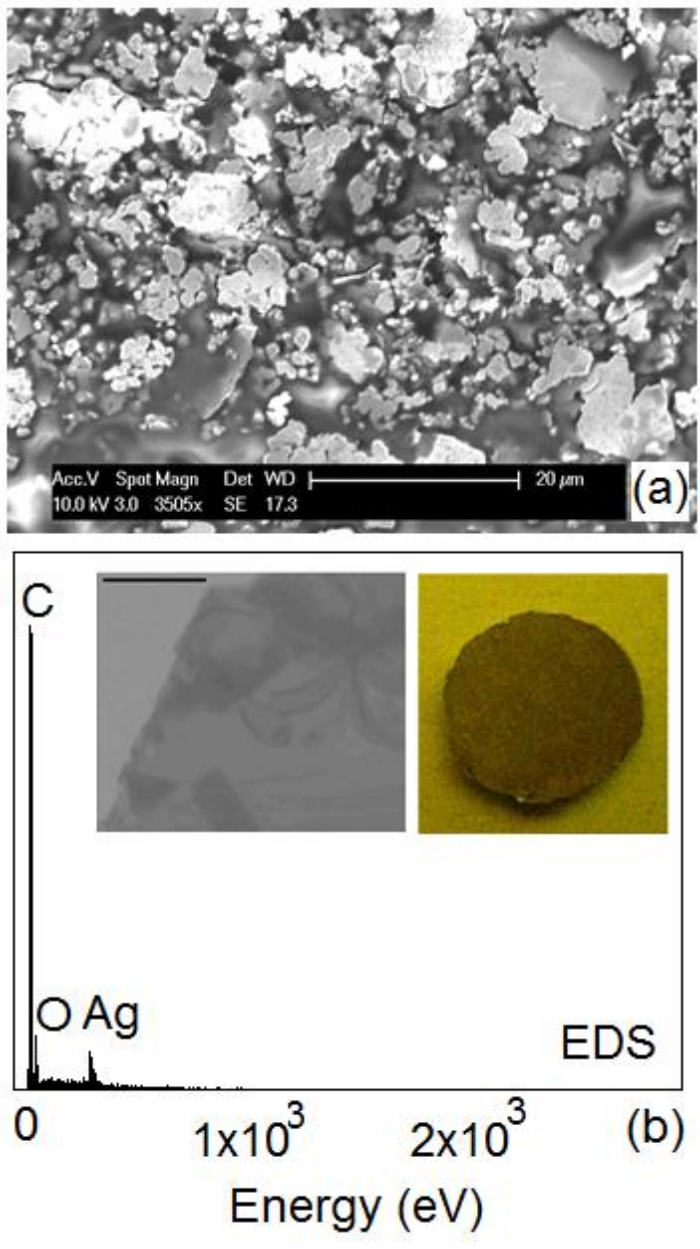

Figure 1 of 3

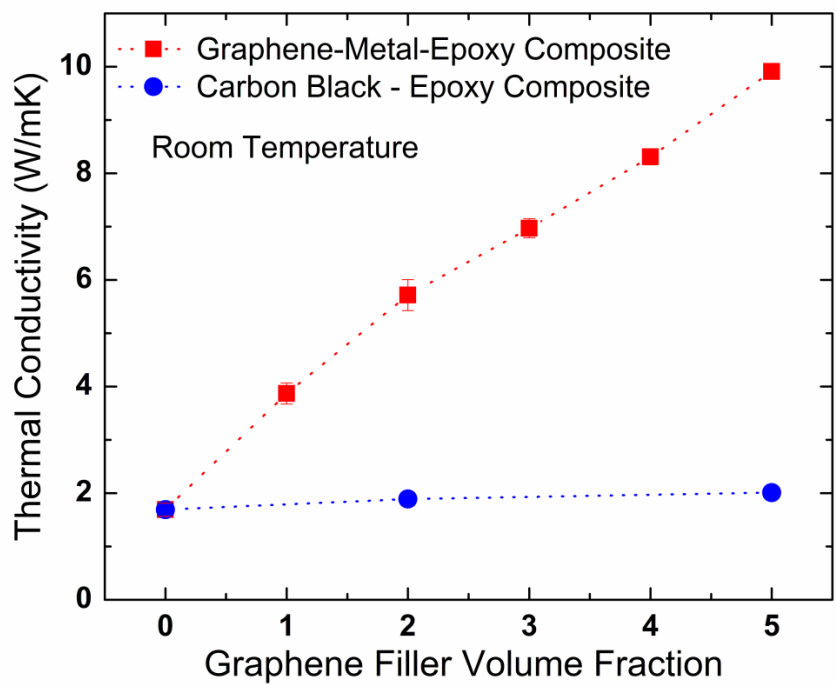

Figure 2 of 3

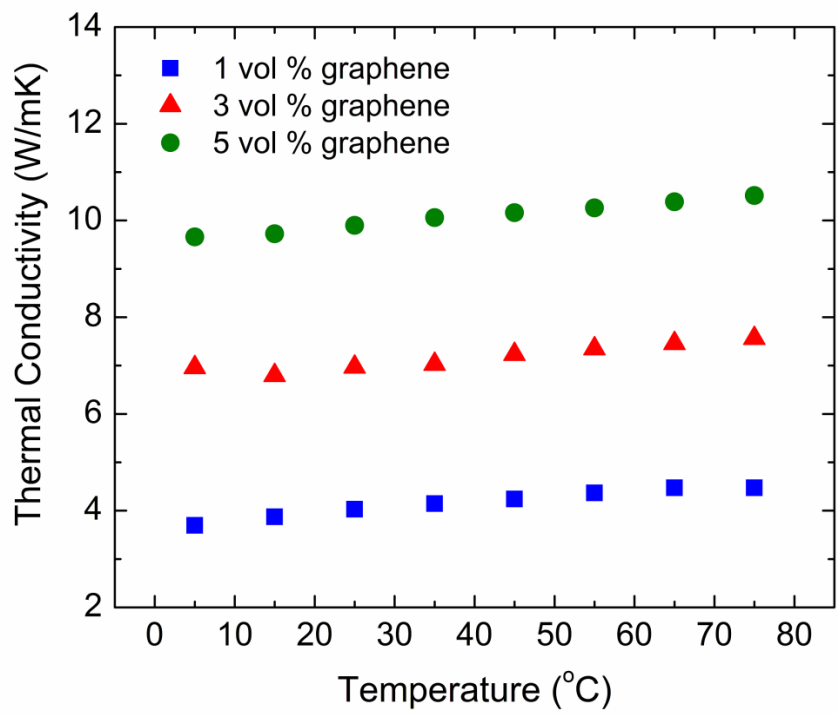

Figure 3 of 3